\documentclass[amsmath,amssymb,superscriptaddress,nobalancelastpage,aps,twocolumn,showpacs]{revtex4-1}

\usepackage{array,multirow,graphicx}
\usepackage{braket}
\usepackage{epstopdf}
\usepackage{graphicx}
\usepackage{siunitx}
\usepackage{nicefrac}
\usepackage[normalem]{ulem}
\usepackage{varioref}
\usepackage{xcolor}
\usepackage{xfrac}
\usepackage{xr-hyper}
\usepackage{hyperref}

\definecolor{bl}{rgb}{0.0,0.2,0.6}
\hypersetup{colorlinks,linkcolor=blue,urlcolor=blue,citecolor=blue}
\newcommand{\mh}[1]{{\textcolor{black}{#1}}}
\newcommand{\mmh}[1]{{\textcolor{black}{#1}}}

\newcommand{\CRO}{Ca$_2$RuO$_4$}

\newcommand{\EF}{$E_\mathrm{F}$}

\begin{document}
	\author{M.~Horio}
	\email{mhorio@issp.u-tokyo.ac.jp}
	\affiliation{Institute for Solid State Physics, The University of Tokyo, Kashiwa, Chiba 277-8581, Japan}
	
	\author{T.~Wada}
	\affiliation{Institute for Solid State Physics, The University of Tokyo, Kashiwa, Chiba 277-8581, Japan}
	
	\author{V.~Granata}
%	\affiliation{CNR-SPIN, I-84084 Fisciano, Salerno, Italy}
%	\affiliation{Dipartimento di Fisica "E.R.~Caianiello", Universit\`{a} di Salerno, I-84084 Fisciano, Salerno, Italy}
	\affiliation{Department of Industrial, Electronic and Mechanical Engineering, Roma Tre University, Via Vito Volterra 62, 00146 Rome, Italy}
	
	\author{R.~Fittipaldi}
	\affiliation{CNR-SPIN, I-84084 Fisciano, Salerno, Italy}
	\affiliation{Dipartimento di Fisica "E.R.~Caianiello", Universit\`{a} di Salerno, I-84084 Fisciano, Salerno, Italy}
	
	\author{A.~Vecchione}
	\affiliation{CNR-SPIN, I-84084 Fisciano, Salerno, Italy}
	\affiliation{Dipartimento di Fisica "E.R.~Caianiello", Universit\`{a} di Salerno, I-84084 Fisciano, Salerno, Italy}
	
	\author{J.~Chang}
	% \email{johan.chang@physik.uzh.ch}
	\affiliation{Physik-Institut, Universit\"{a}t Z\"{u}rich, Winterthurerstrasse 190, CH-8057 Z\"{u}rich, Switzerland}

	\author{I.~Matsuda}
	\affiliation{Institute for Solid State Physics, The University of Tokyo, Kashiwa, Chiba 277-8581, Japan}
	
	%\title{Surface Mott breakdown in \CRO\ induced by ambipolar dopant deposition}
	%\title{Coherent state formation at the surface of Mott-insulating \CRO\ by ambipolar dopant deposition}
	\title{Ambipolar doping-induced surface in-gap state on Mott-insulating \CRO}
	
\begin{abstract}
We report an x-ray photoemission spectroscopy study of \CRO\ surface-dosed with Cs \mh{alkali atoms} and C$_{60}$ \mh{molecules}. Due to its small ionization energy (large electron affinity), deposited Cs atoms (C$_{60}$ molecules) are expected to provide a solid surface with electrons (holes). Upon dosing the dopants to Mott-insulating \CRO, we found a new Ru $3d$ photoemission peak emerging on the lower binding-energy side, suggesting the creation of a core-hole screening channel associated with coherent Ru $4d$ states around the Fermi level. For both the Cs and C$_{60}$ dosing, this change occurred without an appreciable chemical potential jump. The coherent state, therefore, develops within the Mott gap through hybridization with the impurity level of the dopants. The present work highlights the flexibility of Mott-insulator surfaces as a playground for metal-insulator transitions.
\end{abstract}
	
\maketitle

\section{Introduction}
Strongly correlated metals emerging from Mott insulators have attracted broad research interest for these decades~\cite{ImadaRMP1998}. One of the straightforward ways to reach the metallic state is doping carriers away from half-filling. This is realized typically through \mh{chemical} substitution of ions with different valence numbers. Chemical substitutions, however, are inevitably accompanied by a side effect; different ions -- particularly with different valence numbers -- have substantially different radius, and thus substitutions produce strain inside the crystal. Such strain, \mh{known as} chemical pressure, has been a pivotal control parameter for the Mott insulator \CRO~\cite{NakatsujiPRL00,FukazawaJPSJ2000}. \CRO\ belongs to the Ruddlesden-Popper-type layered perovskite and contains RuO$_6$ octahedra as an essential building block. The octahedron undergoes a strong compression along the $c$-axis across 357~K~\cite{FriedtPRB2001}. This first-order transition re-distributes four Ru $4d$ electrons unevenly among the $t_{2g}$ sector, two into $d_{xy}$ and the other two into $d_{xz}/d_{yz}$. Accordingly, the half-filled $d_{xz}/d_{yz}$ orbitals become Mott insulating at 357~K~\cite{NakatsujiJPSJ97,AlexanderPRB1999} due to strong electron correlations arising from Coulomb and Hund's interactions~\cite{GeorgesARCMP2013,SutterNatCom2017}. By applying chemical pressure through isovalent Sr substitution for Ca, the $c$-axis compression is suppressed and so is the Mott transition~\cite{NakatsujiPRL00}. Electron and hole doping has also been attempted by Pr/La substitutions~\cite{FukazawaJPSJ2000} and excess oxygen~\cite{BradenPRB1998,TakenakaNatCommun2017}, respectively, and a metallic state was realized down to the lowest temperature. However, doped carriers are localized and do not directly interact with the Mott-insulating valence bands of \CRO~\cite{RiccoNatCommun18} until the $c$-axis compression is totally suppressed. As such, it \mh{is} unclear which of carrier doping and chemical pressure is the dominant driving force of metallization in these cases.

Surface adsorption of atoms and molecules has been recognized as an effective way to dope a solid surface with carriers. Alkali metals with small ionization energy are often employed as electron dopants~\cite{HossainNature2008,KimScience2014,KimScience2015,YukawaPRB2018,KyungnpjQM2021,HuNatCommun2021} while molecules with large electron affinity like C$_{60}$ are used for hole doping~\cite{LiuAPL2018}. Compared to bulk chemical substitution, the influence on the substrate crystal structure is marginal. \mh{Recent work~\cite{HorioComPhys2023} using angle-resolved photoemission spectroscopy (ARPES)} reported that, with alkali-metal adsorption, a coherent metallic state is formed on the \CRO\ surface. Since the metallic state is composed of a single band~\cite{HorioComPhys2023}, in contrast to the multi-band character of the metallic long-$c$-axis phase of \CRO~\cite{RiccoNatCommun18}, the short-$c$-axis phase is likely retained throughout the metallization. This offers a unique opportunity to study the interaction of doped carriers and the Mott insulating state of \CRO.

Here, we performed an x-ray photoemission spectroscopy (XPS) study of \CRO\ surface-dosed with Cs and C$_{60}$, and monitored electronic-structure evolution caused by ambipolar surface doping. Both with Cs and C$_{60}$ adsorption, a low-energy Ru $3d$ satellite peak emerged in a similar manner, which is a signature of coherent-state formation with significant Ru $4d$ character. These changes occurred without a large jump of the chemical potential across the Mott gap, suggesting that the coherent state lies within the Mott gap. The mechanism of the coherent-state formation is discussed based on the hybridization between the \CRO\ substrate and the dopant impurity level.

\begin{figure*}[ht!]
	\begin{center}
		\includegraphics[width=\textwidth]{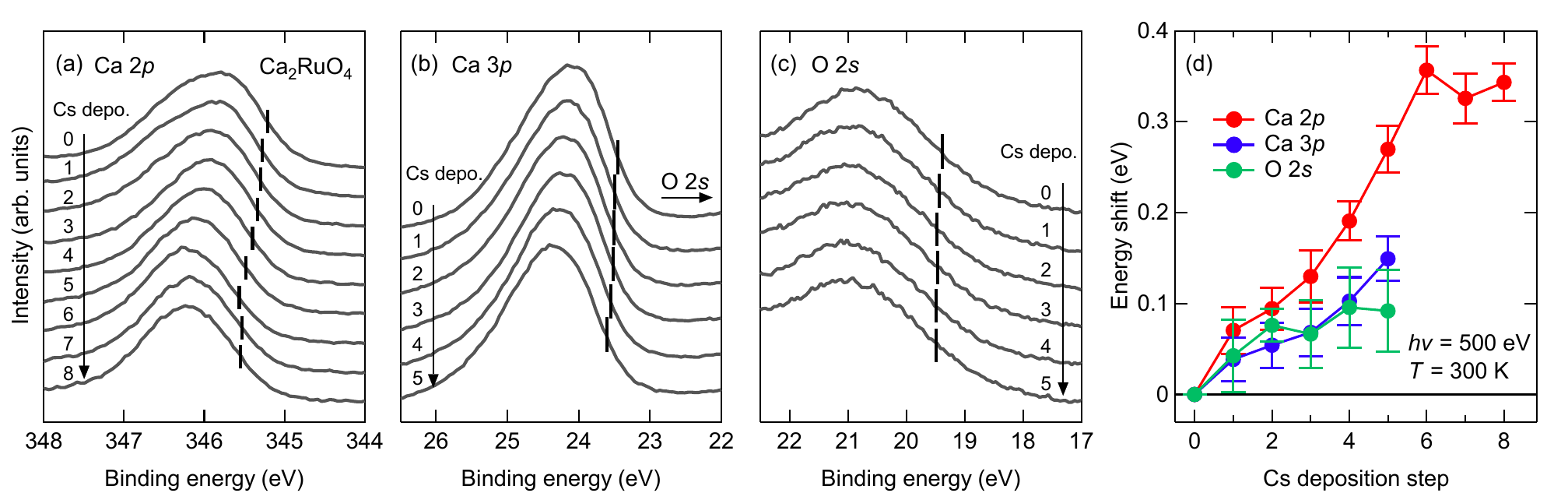}
	\end{center}
	\caption{\textbf{Chemical-potential shift by Cs deposition on \CRO.} (a) Ca~$2p$ spectra recorded at $h\nu = 500$~eV plotted in the order of Cs deposition \mh{steps} from top to bottom. The vertical bars mark the half-maximum position on the lower energy side. After five times of Cs deposition, a new sample was cleaved and deposited with Cs to reach higher dosing (\mh{steps} 6-8). (b),(c) Ca~$3p$ and O~$2s$ spectra of Cs-dosed \CRO, respectively, measured at $h\nu = 500$~eV. All the spectra in (a)--(c) have been normalized to the peak height. (d) The shifts of Ca and O core-level peaks plotted against Cs deposition \mh{steps}. The binominal smoothing filter~\cite{Smoothing} was applied to the spectra before evaluating the half-maximum positions, and the difference from the estimate without smoothing was defined as the error bar.}
	\label{Fig1}
\end{figure*}

\section{Methods}
%sample and general XPS information
Single crystals of \CRO\ were grown by the flux-feeding floating-zone method~\cite{FukazawaPhysB00,NakatsujiJSSC2001}. XPS measurements were carried out at BL07LSU of SPring-8~\cite{YamamotoJSR2014}. Photon energy was varied in the range between 250 and 500~eV with total energy resolution better than 150~meV. \mh{The EPO-TEK E4110 silver epoxy was used for sample mounting and was cured below the structural transition temperature of 357~K.} Samples were cleaved using the top-post method \mmh{in ultra-high vacuum $< 7 \times 10^{-10}$~mbar.}

\begin{figure*}[ht]
	\begin{center}
		\includegraphics[width=\textwidth]{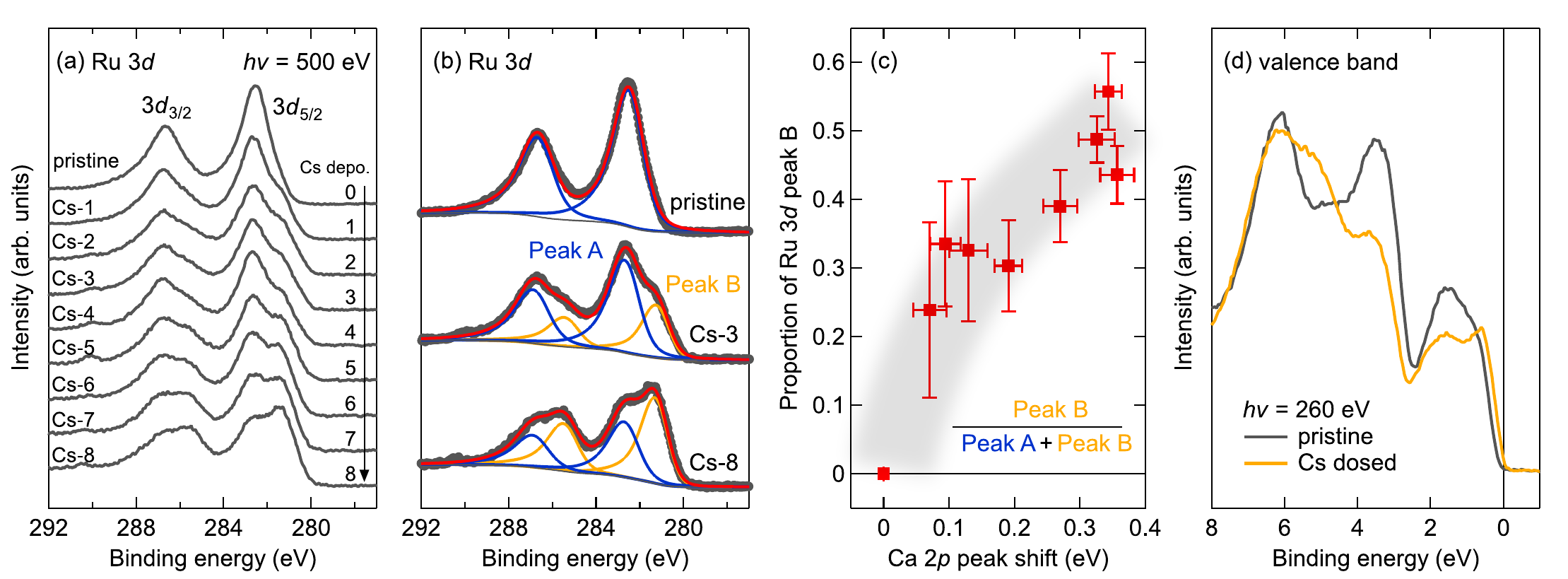}
	\end{center}
	\caption{\textbf{Evolution of the Ru~$3d$ peak upon Cs dosing.} (a) Ru~$3d$ spectra of \CRO\ recorded at $h\nu = 500$~eV and plotted in the order of Cs deposition \mh{steps} from top to bottom. After five times of Cs deposition, a new sample was cleaved and deposited with Cs to reach higher dosing (\mh{steps} 6-8). All the spectra have been normalized to total spectral intensity in the displayed region. (b) Ru~$3d$ spectra of the pristine (Cs-0) and Cs-dosed (Cs-3 and Cs-8) \CRO\ overlaid with fitting curves. See text for details of the fitting procedure. (c) The proportion of peak~B to the overall Ru 3$d$ peak plotted versus Ca~$2p$ peak shift. The vertical error bar has been derived from the deviation of the experimental spectrum from the fitting curve. The gray shade is a guide to the eyes. (d) Valence-band spectra of the pristine and Cs-dosed \CRO\ measured at $h\nu = 250$~eV and normalized to total intensity. The amount of deposited Cs is nominally the same as that for Cs-6 in panel (a).}
	\label{Fig2}
\end{figure*}

%dopant deposition
SAES dispensers were used to deposite Cs onto \CRO, while for C$_{60}$ a Knudsen cell was constructed. Both during the XPS measurements and dopant deposition, samples were kept at room temperature. To ensure pristine surface quality, \mmh{data collection was limited to within 80~minutes after starting the first step of deposition}. For Cs desposition, three samples were measured [\mh{steps} 0-5, \mh{steps} 6-8, and "Cs dosed" in Fig.~\ref{Fig2}(d)], while for C$_{60}$ deposition two samples were measured [data shown separately in Figs.~\ref{Fig3}(a) and (b)]. \mh{One deposition step approximately corresponds to the evaporation of Cs and C$_{60}$ for 30~seconds and 14~minutes with	filament current of 8.3 and 2.5 Ampere, respectively. \mmh{Based on the attenuation of the Ca 2$p$ peak intensity and the known inelastic mean free path (IMFP) of photoelectrons~\cite{TPP,LiuAPL2018}, a single deposition step translates into the layer thickness of $\sim 0.6$~\AA\ ($\sim 1.5$~\AA) for Cs (C$_{60}$), which corresponds to $\sim 0.1$ ($\sim 0.2$) monolayers if the lattice constant of a body-centered-cubic Cs crystal (diameter of a C$_{60}$ molecule) is employed as a unit.}}

\section{Results}
Figure~\ref{Fig1}(a) shows Ca~2$p$ core-level spectra of \CRO\ plotted in the order of Cs deposition \mh{steps}. Referring to the lower binding-energy edge of the spectra as frequently done when discussing XPS peak positions~\cite{InoPRL1997,HarimaPRB2001,FujimoriJSR2002,HorioPRL2018A}, clear shifts toward higher binding energies are observed [see vertical bars in Fig.~\ref{Fig1}(a)]. A similar trend can also be identified for the Ca~3$p$ and O~2$s$ spectra shown in Figs.~\ref{Fig1}(b) and (c), respectively. Inspecting the core-level peak shifts plotted against Cs deposition \mh{steps} [Fig.~\ref{Fig1}(d)], one finds that the shifts for Ca~3$p$ and O~2$s$ are comparable. The shift of core-level binding energy~\cite{Hufner1995,InoPRL1997,HarimaPRB2001,FujimoriJSR2002,HorioPRL2018A} is generally given by
\begin{equation}
\Delta E_\mathrm{B} = \Delta \mu - K\Delta \mathrm{Q} + \Delta V_\mathrm{M} - \Delta E_\mathrm{R},
\label{shift}
\end{equation}
where $\Delta \mu$ is the change in the chemical potential, $\Delta Q$ is the change in the number of valence electrons, $K$ is a constant, $\Delta V_\mathrm{M}$ is the change in the Madelung potential, and $\Delta E_\mathrm{R}$ is the change in the extra-atomic screening of the core-hole potential by conduction electrons and/or dielectric polarization of surrounding media. Comparable shifts observed for the Ca~3$p$ and O~2$s$ core levels indicate that $\Delta V_\mathrm{M}$ is negligibly small because it would shift the core levels of the Ca$^{2+}$ cation and O$^{2-}$ anion in different ways. $\Delta E_\mathrm{R}$ cannot be the main origin of the shifts, either, because introduction of conduction electrons, if any, by Cs deposition would shift the core-level peaks toward lower binding energy, which is opposite to the present observation. Considering that the valences of Ca$^{2+}$ and O$^{2-}$ are fixed ($\Delta Q = 0$), we conclude that the observed shifts in the Ca and O core levels of \CRO\ are largely due to the chemical potential shift $\Delta \mu$.

As compared to Ca~3$p$ and O~2$s$, the Ca~$2p$ peak exhibits larger shifts. \mmh{To address the origin of the deviation, it should be noted that Ca~$2p$ electrons have larger binding energies than those of Ca~3$p$ and O~2$s$, and hence gain smaller kinetic energies by photo-excitation under the fixed photon energy ($h\nu = 500$~eV). This produces a difference in the IMFP of photoelectrons traveling through \CRO; 10~\AA\ for Ca~2$p$ and 21~\AA\ for Ca~3$p$ and O~2$s$, as estimated from the TPP-2M equation~\cite{TPP}. The Ca~$2p$ spectrum thus should reflect surface-specific electronic structure more sensitively than the Ca~$3p$ or O~$2s$ spectrum. In fact, the IMFP of the photo-excited Ca~2$p$ electrons is smaller than the $c$-axis lattice constant of \CRO\ at room temperature (11.96~\AA)~\cite{AlexanderPRB1999}. Larger shifts for the Ca~$2p$ peak, therefore, suggest that the chemical-potential shift is confined at the surface of \CRO. This is consistent with the adsorption, rather than intercalation, of alkali metal atoms.}

Turning to the Ru~$3d$ core-level spectra, one finds a single set of spin-orbit-split peaks on the pristine surface [Fig.~\ref{Fig2}(a)]. Upon Cs deposition, a new component develops on the lower binding energy side of both the $3d_{5/2}$ and $3d_{3/2}$ peaks [Fig.~\ref{Fig2}(a)]. To capture this change in a quantitative manner, we fitted the spectra to two sets of spin-orbit-split peaks plus a Shirley-type background. Peaks were modeled by the Mahan line shape
\begin{equation}
	\frac{1}{\mathrm{\Gamma}(\alpha)} \frac{e^{-(E_B-E_0)/\xi}}{\left| (E_B-E_0)/\xi \right| ^{1-\alpha}} \Theta(E_B-E_0)
	\label{Mahan}
\end{equation}
convolved with a Voigt function, where $\alpha$ is asymmetry, $\xi$ is a cut-off parameter, and $E_0$ is the peak position. Peak separation and intensity ratio between the spin-orbit-split 3$d_\mathrm{5/2}$ and 3$d_\mathrm{3/2}$ peaks were fixed at 4.17~eV and 3:2, respectively~\cite{RuXPS_2015}. Gaussian width for the Voigt function and $\xi$ were assumed identical for all the peaks, and within the spin-orbit-split pair an identical value was used for $\alpha$. We first fitted the Ru~$3d$ spectrum after the seventh Cs deposition \mh{step} (Cs-7), where both the higher and lower energy components are clearly identified. Then, allowing only the peak intensity and positions to vary, the rest of the Ru $3d$ spectra were fitted. The results of fitting are displayed in Fig.~\ref{Fig2}(b) for the pristine and some of the Cs-dosed surfaces. The spectrum after three times of Cs deposition (Cs-3) is reproduced by the combination of peak~A -- with the same spectral lineshape as the pristine-surface case -- and peak~B at a lower binding energy, demonstrating the transition from single to double components by Cs deposition. Spectral changes upon further deposition (Cs-8) can be reproduced simply by the enhancement of peak~B relative to peak~A. %Since the binding energy is comparable between Ru~$3d$ and Ca~$2p$, one can regard the changes of both core levels as occurring around the same depth level.
The proportion of peak~B to the overall Ru 3$d$ peak is plotted in Fig.~\ref{Fig2}(c) as a function of the Ca~$2p$ shift, i.e., chemical potential shift. The proportion evolves accordingly to the chemical potential shift and exceeds 0.5 at the largest. Note that the binding energies of Ru~3$d$ and Ca~2$p$ are relatively close, and hence the probing depth are comparable between the two cases.

\begin{figure}[ht]
	\begin{center}
		\includegraphics[width=0.47\textwidth]{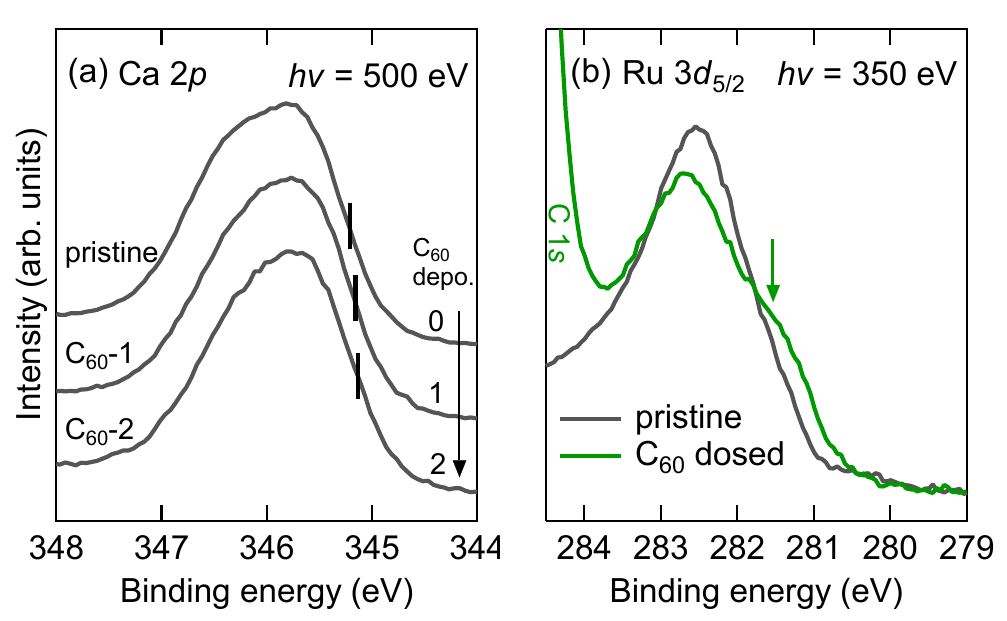}
	\end{center}
	\caption{\textbf{Effect of C$_{60}$ dosing on the core-level spectra of \CRO.} (a) Ca~$2p$ spectra recorded at $h\nu = 500$~eV plotted in the order of C$_{60}$ deposition \mh{steps} from top to bottom. The vertical bars mark the half-maximum position on the lower energy side. (b) Comparison of the Ru~$3d$ spectra of the pristine and C$_{60}$-dosed \CRO\ measured at $h\nu = 350$~eV. The amount of dosed C$_{60}$ is nominally the same as that for C$_{60}$-1 in (a). The green arrow marks the additional lower-energy component emerging after C$_{60}$ deposition. The steep increase of intensity above 284~eV after C$_{60}$ dosing is due to the C 1$s$ peak of C$_{60}$.} 
	\label{Fig3}
\end{figure}

The emergence and growth of peak~B occur along with a clear change in the valence-band spectrum. As shown in Fig.~\ref{Fig2}(d), pristine \CRO\ is in a Mott-insulating state and, therefore, the valence-band spectrum is gapped with the lower Hubbard band positioned around 1.7~eV below the Fermi level (\EF)~\cite{SutterNatCom2017}. After Cs deposition of nominally the same amount as that for Cs-6 in Fig.~\ref{Fig2}(a), spectral weight of the lower Hubbard band is strongly suppressed, and instead finite spectral weight accumulates near \EF, consistent with the previous ARPES study~\cite{HorioComPhys2023}. Peak~B in the Ru~$3d$ spectrum could thus be associated with the low-energy states around \EF. %That is, a new final state becomes available after Cs dosing where Ru~$3d$ core holes are screened by newly created Ru~$4d$ coherent states. Such core-hole screening by Ru~$4d$ conduction electrons have been ubiquitously observed in the previous XPS studies of strongly correlated metallic ruthenium oxides~\cite{Cox1983,KimASS1997,KimPRL2004,GuoPRB2010,PanaccioneNJP2011}.

Having overviewed the effect of Cs dosing to \CRO, let us now examine the C$_{60}$-dosed case. As shown in Fig.~\ref{Fig3}(a), the Ca~$2p$ peak is shifted towards the lower binding-energy side upon C$_{60}$ deposition. This is opposite to the change caused by Cs dosing [Fig.~\ref{Fig1}(a)]. Nevertheless, the Ru~$3d$ spectrum shows a similar change, that is, the emergence of another peak on the lower energy side as indicated by the arrow in Fig.~\ref{Fig3}(b). %The distance between the two peaks, determined from second derivative of the spectrum, is 1.25~eV, which is close to 1.35~eV evaluated by fitting for the Cs-dosed case, suggesting the same mechanism lying behind the peaks. The Ru~$3d$ spectra from the pristine, Cs-dosed, and C$_{60}$-dosed surfaces can be directly compared in Fig.~\ref{Fig4}(a).

\begin{figure}[ht]
	\begin{center}
		\includegraphics[width=0.48\textwidth]{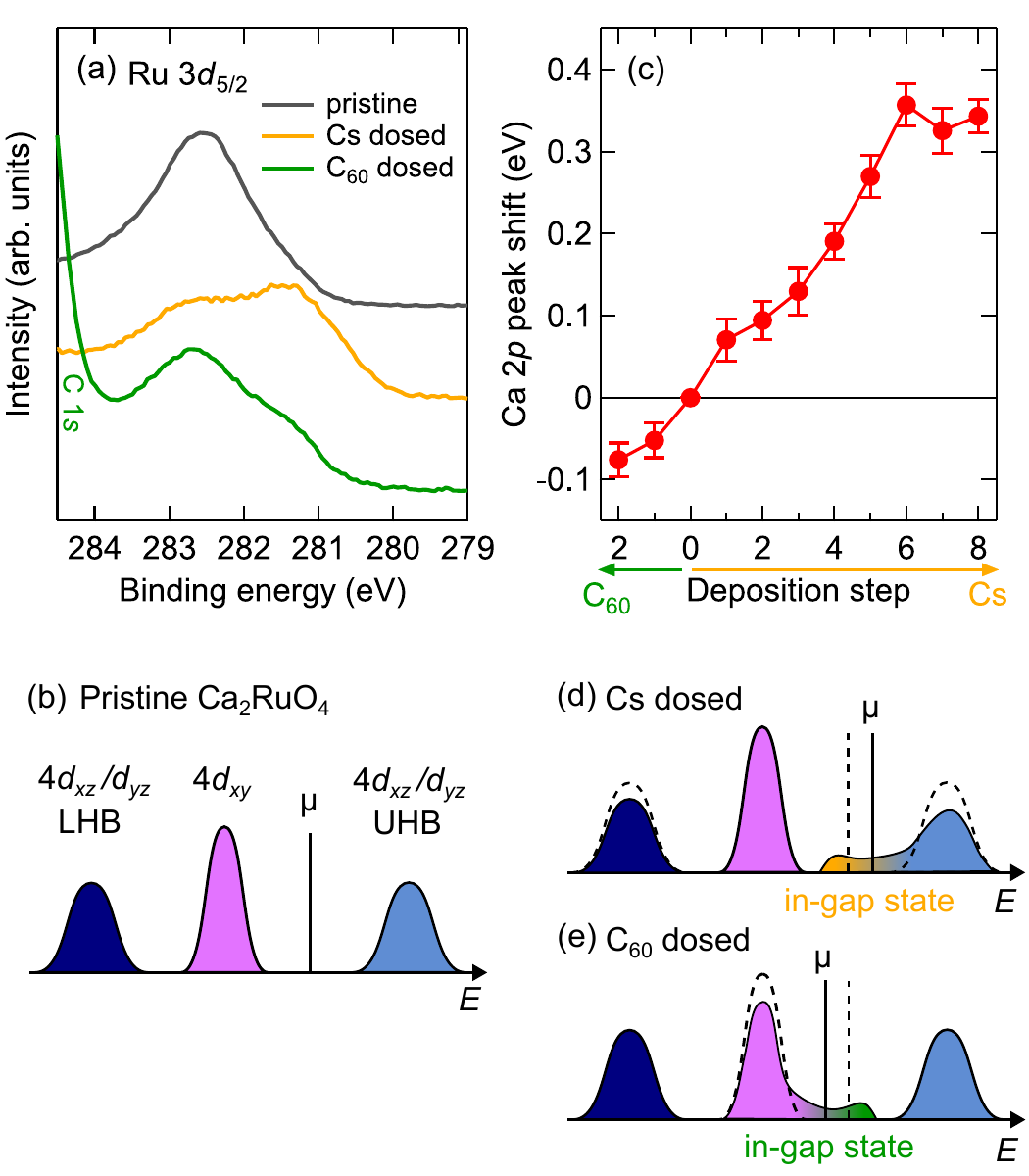}
	\end{center}
	\caption{\textbf{Smooth evolution of the coherent in-gap state.} (a) Ru $3d$ spectra of the pristine, Cs-dosed, and C$_{60}$-dosed \CRO. The spectra of the pristine and Cs-dosed ones have been taken from Fig.~\ref{Fig2}(a), and the C$_{60}$-dosed one from Fig.~\ref{Fig3}(b). \mh{(b) Schematic energy diagram for the pristine \CRO. UHB and LHB denote the upper and lower Hubbard bands, respectively.} \mh{(c)} The shift of the Ca $2p$ peak, which represents the chemical potential shift, as a function of Cs or C$_{60}$ dosing. No abrupt jump is observed in either direction. \mh{(d),(e) Schematic energy diagrams of in-gap state formation after Cs and C$_{60}$ dosing, respectively.}} 
	\label{Fig4}
\end{figure}

\section{Discussion}
It is of great importance to discuss and clarify the origin of the low-energy peak in the Ru~$3d$ spectra that emerges after dopant deposition. Since lower binding energy of core-level spectra is in general associated with more negatively charged states, one might suspect that the new peak arose merely from the lower valence state of Ru (e.g. Ru$^{3+}$). Indeed, previous XPS studies on transition-metal oxides found such a peak after surface electron doping by Cs adsorption~\cite{YukawaPRB2018}. However, the proportion of the peak from different valence should be at most $\sim 0.15$, which is the limit of electron doping by surface-adsorbed alkali metals generally observed for transition-metal oxides~\cite{HossainNature2008,KimScience2014,YukawaPRB2018,KyungnpjQM2021,HuNatCommun2021}. %even though the exact value depends on the sample work function~\cite{JungJESRP2021}. 
In the present Cs-dosed case, the proportion of the new peak exceeds 0.5 at maximum, which is unrealistically large if it solely originated from the electron-doped different valence state of Ru. Moreover, similar peak was found after dosing C$_{60}$ [Fig.~\ref{Fig3}(b)], which is known for large electron affinity and thus as a hole dopant~\cite{LiuAPL2018}. In fact, the downward chemical-potential shift observed for C$_{60}$-dosed \CRO\ [Fig.~\ref{Fig3}(a)] is not compatible with electron doping. Nevertheless, the peak positions of the Cs- and C$_{60}$-dosed cases are quantitatively comparable [Fig.~\ref{Fig4}(a)], suggesting a common origin of the two-peak structure. %Instead of chemically inequivalent species, multiple photoemission final states common to both dosing cases are thus inferred.

The two-peak Ru $3d$ structure has been ubiquitous for metallic ruthenium oxides while insulators display only a single peak~\cite{Cox1983,KimPRL2004,PanaccioneNJP2011,GuoPRB2010}. Several mechanisms involving core-hole screening have been proposed. One ascribes the lower energy peak to a final state where core holes created by photoemission are non-locally screened by charge transfer from neighboring Ru atoms~\cite{Okada2002,VeenendaalPRB2006}. The peak intensity depends on the amount of holes in the Ru $4d_{xy}$ orbital as it hybridizes strongly with in-plane oxygen orbitals and hence can be responsible for the charge transfer. This scenario is consistent with a single Ru $3d$ peak for pristine \CRO\ with the fully occupied $4d_{xy}$ orbital. However, gradual evolution of the lower-energy peak along with the continuous upward shift of the chemical potential by alkali-metal adsorption is incompatible with hole creation in the $4d_{xy}$ orbital whose maximum is positioned at $\sim 0.3$~eV below \EF~\cite{SutterNatCom2017}. On the other hand, Kim \textit{et al.}~\cite{KimPRL2004} proposed -- through dynamical mean field theory  approach --  an alternative mechanism which is free from orbital character assumptions. Here, coherent quasi-particles with Ru $4d$ character on the Fermi surface was shown to act as a source of core-hole screening. This scenario, associating the lower-energy peak directly with a spectral weight at \EF, has successfully reproduced the core-level evolution of ruthenates upon metal-insulator transitions~\cite{KimPRL2004}. The mechanism is applicable irrespective of band filling and hence to both the Cs- and C$_{60}$-dosed cases. The observed valence-band evolution [Fig.~\ref{Fig2}(d)] is also in line with this mechanism. We hence conclude that the low-energy Ru $3d$ peak emerging upon Cs and C$_{60}$ adsorption on \CRO\ is mainly due to the development of coherent quasi-particles near \EF.

When a Mott insulator is doped with electrons and holes, they are expected to enter the upper and lower Hubbard bands, respectively, and form coherent quasi-particles to reach a metallic state. At the boundary of the electron and hole doping, an abrupt chemical-potential jump occurs between the bottom of the upper Hubbard band and the top of the lower Hubbard band~\cite{IkedaPRB2010,HuNatCommun2021}. For \CRO, \mh{a straightforward expectation is that the chemical potential jumps between the bottom of the upper Hubbard band with the Ru 4$d_{xz}/d_{yz}$ character and the top of the Ru 4$d_{xy}$ band residing within the Mott gap [Fig.~\ref{Fig4}(b)]. However,} we did not observe a clear jump [Fig.~\ref{Fig4}(c)] despite the signature of coherent state formation. The absence of a chemical potential jump suggests that carriers provided by Cs or C$_{60}$ are bound on the dopant sites and form an impurity level rather than injected into the \mh{Ru 4$d$} bands of the substrate \CRO. Such a bound state could weakly hybridize with substrate orbitals, given enough spatial overlap and symmetry compatibility~\cite{ZhangPRB2016}. The overlap of wave function and hence orbital hybridization should increase accordingly with surface coverage and \mh{smoothly evolve into} a hybrid, coherent in-gap state \mh{[Figs.~\ref{Fig4}(d) and (e)].} This consideration reconciles the existence of coherent low-energy state and the absence of chemical-potential jump. \mh{The hybrid in-gap state of the Ru 4$d_{xz}/d_{yz}$ and alkali $s$ orbitals [Fig.~\ref{Fig4}(d)] has indeed been identified by ARPES~\cite{HorioComPhys2023}.} \mmh{Analogously, the simplest mechanism of the in-gap state formation on the C$_{60}$-dosed side would be to involve the $d_{xy}$ orbital, whose energy is the highest among the occupied states [Fig.~\ref{Fig4}(e)]. On the other hand, the possibility of additional contributions from the $d_{xz}/d_{yz}$ orbitals, through non-trivial modifications of the Hubbard bands, are not excluded at present. The exact orbital character of this in-gap state could be a target of future ARPES investigations.}
	
%ARPES investigations into the C$_{60}$-induced state is an interesting future work, which should give a clue about whether the hybrid state is simply derived from the $d_{xy}$ orbital as illustrated in Fig.~\ref{Fig4}(e), or the $d_{xz}/d_{yz}$ orbitals are also involved in a non-trivial manner.

For \CRO, a small amount ($\sim$ 0.03 per Ru atom) of electrons doped by Pr/La substitution is known to localize without causing appreciable changes to the electronic structure~\cite{RiccoNatCommun18}. A metallic state can be realized by increasing Pr/La concentration, but is accompanied by a prominent $c$-axis elongation~\cite{FukazawaJPSJ2000,CaoPRB2000}. This implies a significant role played by chemical pressure in this metallization. As for surface dosing, the increase of carrier doping (or orbital hybridization) is possible without significantly modifying the crystal structure. This characteristics allowed us to surpass the localization regime while keeping the lattice intact and to investigate direct interactions between the doped carriers and Mott-insulating substrate albeit in a surface specific manner. As such, surface dopant adsorption could lead us to an unexplored region in the parameter space and expand the potential of Mott insulators as a source of strongly correlated metals.\\

\section{Conclusions}
To summarize, we performed the XPS measurements of the Mott insulator \CRO\ surface-dosed stepwise with Cs and C$_{60}$. Upon dopant deposition, Ca and O core levels shifted in the same direction by comparable amounts, suggesting the shift of chemical potential. The direction of the chemical potential shift was found to be opposite between the Cs- and C$_{60}$-dosing cases, reflecting the opposite signs of doped carriers. Nevertheless, the Ru~3$d$ spectrum showed similar changes, that is, the emergence of another peak on the lower energy side, which is associated with a photoemission final state where core holes are efficiently screened by coherent states near \EF. Since this change in Ru~3$d$ was observed on both the electron- (Cs) and hole- (C$_{60}$) doped sides without abrupt chemical-potential jump, the coherent state should be created within the Mott gap through hybridization with the impurity level of the dopants. Such smooth coherent-state formation is in contrast to the drastic metal-insulator transition with the structural modification induced by bulk chemical substitutions. Surface adsorption of dopants could thus offer a path toward the metallization of Mott insulators, which is distinct but complementary to bulk carrier doping.

\section{Acknowledgements}
Fruitful discussion with F.~Forte and M.~Cuoco is greatfully acknowledged. This work was supported by JSPS KAKENHI Grant Numbers JP21K13872 and JP25K00218. \mmh{R.F. and A.V. acknowledge support by the Italian Ministry of Foreign Affairs and International Cooperation, grant KR23GR06.} The measurements were performed with the approval of the Japan Synchrotron Radiation Research Institute (JASRI) under Proposal No.~2021B7411.\\

%\bibliography{CRO}

\end{document}